\documentclass[twocolumn,showpacs,floatfix,aps,pre]{revtex4}
\usepackage{graphicx}% Include figure files
\usepackage{dcolumn}% Align table columns on decimal point
\usepackage{bm}% bold math
\usepackage{color} %color

\newcommand{\tr}{t_{\rm r}}
\newcommand{\ep}{\varepsilon}

\begin{document}

\title{
Scaling in activated escape of underdamped
systems}

\author{M.I. Dykman}
\affiliation{Department of Physics and Astronomy, Michigan State
University, East Lansing, MI 48824, USA}
\author{I.B. Schwartz}
\affiliation{Plasma Physics Division, Code 6792, Naval Research
Laboratory, Washington DC 20375}
\author{M. Shapiro}
\affiliation{Department of Mathematics, Michigan State
University, East Lansing, MI 48824, USA}

\noindent
\begin{abstract}
Noise-induced escape from a metastable state of a dynamical system is
studied close to a saddle-node bifurcation point, but in the region
where the system remains underdamped. The activation energy of escape
scales as a power of the distance to the bifurcation point. We find
two types of scaling and the corresponding critical exponents.
\end{abstract}


\maketitle

\section{Introduction}

Noise-induced escape from a metastable state underlies many important
phenomena in nature and technology. It is often investigated
experimentally for systems close to a bifurcation point where a
metastable state disappears. Examples include Josephson junctions
\cite{Fulton74,Devoret87,Siddiqi04}, nanomagnets
\cite{Wernsdorfer97,Koch00,Ralph02}, and mechanical nano- and
microresonators \cite{Cleland05,Mohanty_APL05,Chan05}. Studying escape
in this regime is advantageous in two respects. First, near a
bifurcation point the activation barrier is small, so that the escape
rate can be measured in a conceivable time even for low temperature or
small noise intensity. Second, the dynamics near a bifurcation point
displays system-independent features, including the onset of one or
several soft modes. In turn, this leads to scaling of the escape rate
with the control parameter of the system.

We will discuss escape for system near a generic saddle-node
bifurcation, which is one of the most common types of
bifurcations. Here, for very small values of the control parameter
$\eta$ counted off from the bifurcation value $\eta_B=0$, the
metastable and unstable states of the system are close to each other
in phase space. The motion along the axis that connects them is
slow. This is easy to understand when metastable and unstable states
are a minimum and a saddle point of the potential energy of a
particle. The curvature of the potential has opposite signs at these
points. Therefore the potential locally flattens out as $\eta$
approaches zero. In agreement with this picture, for very small $\eta$
the system dynamics is controlled by a single soft mode $q$, which is
overdamped.

In this paper we consider systems with very small damping. Their
motion may become underdamped even where the parameter $\eta$ is still
close, although not too close to the bifurcation value. The dynamics
of the soft mode is characterized by the vibrational frequency at the
metastable state $\omega_a$. It is much less than other frequencies,
and $\omega_a\to 0$ for $\eta\to 0$. Yet $\omega_a$ may exceed the
relaxation rate, which is characterized by an independent small
parameter $\varepsilon$. Such a situation often occurs in underdamped
Josephson junctions, for example.

Noise-induced escape near a bifurcation point is well understood in
the limit $\eta\to 0$, where the motion is overdamped
\cite{Dykman80,Graham_bifurc,Dykman_Ross94,Tretiakov03}. Here, under
fairly general assumptions the activation energy of escape scales as
$\eta^{3/2}$. This agrees with the scaling of the mean-field
free-energy barrier near a termination point of the metastable state
\cite{LL_Stat}. As we show, such scaling holds also for underdamped
systems further away from $\eta=0$ provided the metastable and saddle
states are still close to each other. This can be understood for a
Brownian particle in a potential well. Here, the activation energy of
escape is given by the height of the potential barrier, independent of
friction; friction only affects the prefactor in the escape rate
\cite{Kramers}. Therefore the scaling is the same in the under- and
overdamped regimes. The scaling $\eta^{3/2}$ was discussed already in
the early work on Josephson junctions \cite{Kurkijarvi72} and later
for underdamped driven oscillators \cite{Dyakonov86} and magnets
\cite{Victora89}.

As we show, besides the $\eta^{3/2}$ scaling, underdamped systems
may display a different scaling near, but not too close to the
bifurcation point. This is because the motion slowing down is not
necessarily related to the stable and unstable state approaching
each other. The slowing down may have a different, nonlocal
origin. It can be thought of as resulting from flattening, or an
overall decrease in magnitude of the effective potential of the
system in a broad range of the generalized coordinate. We provide
a theory of escape for this nonlocal case and find the
corresponding critical exponent. The results are applied to a
specific system, which provides an example of this scaling
behavior and has attracted much attention recently
\cite{Siddiqi04,Cleland05,Mohanty_APL05,Chan05}.

In Sec.~II we discuss the model of a fluctuating underdamped
system. In Sec.~III we describe two scenarios that lead to the motion
slowing down and ultimately to the saddle-node bifurcation, in the
presence of friction, as the control parameter $\eta\to 0$. The effect
of friction is described in Sec.~IV. Energy diffusion is considered in
Sec.~V. In Sec.~VI we obtain the critical
exponents for the activation energy of escape, which is the central
result of the paper. In Sec.~VII the theory is applied to a nonlinear
oscillator driven by a strong resonant field and compared with the
previous results for this model \cite{Dyakonov86,DK79}. Sec.~VIII
contains concluding remarks.

\section{The model}

Close to a saddle-node bifurcation point one of the motions in the
system becomes slow compared to other motions. Quite generally it is
described by a Langevin equation
\begin{eqnarray}
\label{eom}
&&\dot q = \partial_pH(p,q) - \varepsilon v^{(q)}(p,q)+f^{(q)}(t),\nonumber\\
&&\dot p = -\partial_qH(p,q) - \varepsilon v^{(p)}(p,q)+f^{(p)}(t).
\end{eqnarray}
Here, $q$ and $p$ are the coordinate and momentum of the slow motion,
and $H$ is its effective Hamiltonian. The terms $\varepsilon
v^{(q,p})$ and $f^{(p,q)}$ describe friction and noise,
respectively. We have assumed that the motion is slow on the scale of
the correlation time of the reservoir coupling to which leads to
friction and noise. Therefore we disregard retardation in the friction
force.  In the same approximation we disregard the noise correlation
time and set the noise to be $\delta$-correlated in time,
\begin{equation}
\label{white_noise}
\langle f^{(i)}(t)f^{(j)}(t')\rangle = 2D_{ij}(p,q)\delta(t-t'),
\end{equation}
where $i,j$ enumerate the $q$ and $p$ components of noise. The
diffusion matrix is symmetric, $D_{qp}=D_{pq}$, and generally $D_{ij}$
depends on the dynamical variables $q,p$.  The noise intensity
$\propto \max D_{ij}$ will be the smallest parameter of the theory.

We assume that the friction coefficient $\varepsilon$ is small and
that the friction force is non-Hamiltonian. The latter requires that
\begin{equation}
\label{div_f}
{\bm\nabla}{\bf v} \equiv \partial_qv^{(q)}+\partial_p v^{(p)}
\end{equation}
is not equal to zero.
%In fact, we will require that ${\bm\nabla}{\bf
%v}\neq 0$ in the entire region of the phase space of interest.
The part of the force $(v^{(q)}, v^{(p)})$ for which ${\bm\nabla}{\bf
v}=0$ can be incorporated into the Hamiltonian, leading to its small
renormalization $\propto \varepsilon$.  This concerns, in particular,
$q$-independent terms in $v^{(q)}$ and $p$-independent terms in
$v^{(p)}$.

The Hamiltonian $H(p,q)\equiv H(p,q;\eta)$ depends on the control
parameter $\eta$ which characterizes the distance to the bifurcation
point. This point is located at $\eta=0$, for $\varepsilon=0$. We
require that (i) $H(p,q;\eta)$ be analytic in $\eta$ in a neighborhood
of $\eta=0$, and (ii) $\partial_{\eta} H\not\equiv 0$ for $\eta=0$.
These two conditions determine the order of the Hamiltonian $H$ with
respect to $\eta$ uniquely for the different functional forms of $H$
used below, i.e., if both $\eta_1$ and $\eta_2$ meet them, then
$\eta_2=O(\eta_1)$.

The physical picture of motion leading to escape is simple. The system
is initially prepared at some point $(q,p)$ in the basin of attraction
to the metastable attractor (focus) $(q_a,p_a)$. Over the relaxation
time $\tr \propto \varepsilon^{-1}$, the system approaches the
attractor. Noise leads to small fluctuations about the attractor, but
for small noise intensity the system stays close to it for a time that
largely exceeds $\tr$. Ultimately there happens a large fluctuation
that carries the system over the basin boundary leading to escape from
the metastable state.

\subsection{Brownian particle near
a saddle-node bifurcation point: standard analysis}

The probability $W$ of noise-induced escape from a metastable state is
a complicated function of the dynamical parameters. However, it
displays a universal behavior in the vicinity of a bifurcation
point. This behavior has been well understood in two limiting
cases. The best known case is a Brownian particle in a potential well,
which escapes from the well due to thermal fluctuations.  The motion
of a Brownian particle of mass $m=1$ in a potential $U(q)$ is
described by the equation (\ref{eom}) with
\begin{eqnarray}
\label{equilibrium}
&&H(p,q)=\frac{1}{2}p^2 + U(q), \qquad f^{(q)}=0,\nonumber\\
&&v^{(q)}=0,\qquad v^{(p)}= p,
\end{eqnarray}
and with $f^{(p)}(t)$ being white noise of intensity $D=2\varepsilon
kT$.

The dynamics of a Brownian particle is special in several
respects. First, the value of the momentum in a stationary state is
independent of the parameters and is equal to zero. Second,
independent of the friction coefficient, the stable and unstable
states of the system $q_a$ and $q_s$ are the local minimum and maximum
of $U(q)$. Third, as a consequence of the form of the noise, the
escape rate has a simple form
\begin{equation}
\label{thermal}
W={\rm const}\times e^{-\Delta U/kT},
\qquad \Delta U=U(q_s)-U(q_a).
\end{equation}
The exponential factor here is just the ratio of the Boltzmann
factors in the populations of states with energies of the
saddle point and the attractor. It is {\it independent} of the
friction coefficient. In contrast, the prefactor in $W$ depends on
$\ep$ \cite{Kramers}; however, in this work we will be interested only in the
exponent in $W$.

The potential $U(q)$ depends on the control parameter $\eta$. The
standard analysis of this dependence goes as follows.  When $\eta$ is
close to the critical value $\eta=0$, the stationary states are close
to each other and ultimately merge for $\eta=0$; we set $q_a=q_s=0$
for $\eta=0$. For small $\eta$ the potential is a cubic parabola,
$U(q) \approx U_{\rm cub}(q)$,
\begin{equation}
\label{cubic}
U_{\rm cub}(q)=-q^3/3 + \eta q
\end{equation}
locally in $q$ (which has been appropriately scaled). The stable and
unstable states are $q_a=-\eta^{1/2},\, q_s=\eta^{1/2}$. Both the
distance $q_s-q_a$ and the frequency of vibrations about the potential
minimum in the absence of friction $\omega_a=(4\eta)^{1/4}$ decrease
with decreasing $\eta$, as expected.

Eq.~(\ref{cubic}) leads to the scaling of the activation barrier with
$\eta$ of the form
$\Delta U= (4/3)\eta^{3/2}$.
We emphasize that this scaling holds independent of friction, for a
Brownian particle. Because the friction coefficient $\ep$ is small,
the approximation (\ref{cubic}) applies both where the system is
underdamped, with $\omega_a\gg \ep$, or underdamped, with $\omega_a\ll
\ep$. The critical value of the bifurcation parameter $\eta$ is also
independent of $\ep$.

\subsection{Close vicinity of the bifurcation point}

Irrespective of the friction coefficient, very close to the
bifurcation point the dynamics of the system is overdamped
\cite{Guckenheimer}. It is described by a soft mode $Q$, which
is the distance between the saddle and the attractor. The Langevin
equation for $Q$ after proper scaling and changing to a slow time
$\tau$ can be written as
\begin{equation}
\label{overdamped}
\frac{dQ}{d\tau} = Q^2-\eta + f^{(Q)}(\tau),
\end{equation}
For $\eta > 0$ the system has a stable and unstable states $Q=\mp
\eta^{1/2}$, which merge together for $\eta = 0$.

Because the motion near the bifurcation point is slow, with
characteristic time $\eta^{-1/2}$, and occurs in a narrow range
$|Q|\lesssim\eta^{1/2}$, the noise $f^{(Q)}(\tau)$ can be considered
white, with a $Q$-independent intensity $D$. Then the escape rate is
\cite{Dykman80}
\begin{equation}
\label{escape_overdamped}
W= \frac{\eta^{1/2}}{\pi}e^{-R/D},\qquad R=\frac{4}{3}\eta^{3/2}.
\end{equation}
Eqs.~(\ref{overdamped}), (\ref{escape_overdamped}) can be understood
as the limit of Brownian motion for $\ep\gg \eta^{1/4}$, where the
Brownian motion becomes overdamped. Eq.~(\ref{overdamped}) follows
from Eqs.~(\ref{eom}), (\ref{equilibrium}) in this limit, with $Q=q$
and $\tau=\ep t$. The activation energy of escape $R$
(\ref{escape_overdamped}) scales with $\eta$ in the same way as the
barrier height $\Delta U$ for the corresponding Brownian particle.

We emphasize that the applicability of Eq.~(\ref{overdamped}) is not
at all limited to systems close to thermal equilibrium. It applies
essentially to any weakly fluctuating dynamical system close to a
saddle-node bifurcation point, see \cite{Tretiakov03} and references
therein.  The scaling $R\propto \eta^{3/2}$ has been considered a
benchmark of activated escape near a saddle-node bifurcation point. It
has been used very broadly, in particular as a tool for measuring
critical current of Josephson junctions \cite{Fulton74,Devoret87}.

\section{Saddle-node bifurcation in underdamped nonequilibrium systems}

We will consider the general case of a system away from thermal
equilibrium, which is described by Eq.~(\ref{eom}) with small friction
coefficient $\ep$. Our major assumption is that, as we decrease the
control parameter of the Hamiltonian $\eta$ from a positive value
$\sim O(1)$ while keeping $\ep =$const, the system will experience a
saddle-node bifurcation. Because the system is nonequilibrium, the
bifurcation value $\eta_B$ may depend on the friction parameter
$\varepsilon$. For convenience, in what follows we do not set
$\eta_B=0$, but we assume
\[ |\eta_B|\ll 1, \qquad \lim_{\ep \to 0}\eta_B=0.\]

For small $\ep$ the system is underdamped for $\eta_B\ll \eta\ll
1$. Still very close to $\eta_B$ it invariably becomes overdamped and
is described by Eq.~(\ref{overdamped}).  We want to find the {\it
forms of the Hamiltonian} and the corresponding phase portraits of the
Hamiltonian trajectories that will be compatible with this
scenario of the saddle-node bifurcation.

By assumption, the bifurcation results from the change of the
Hamiltonian dynamics, not from the special structure of the friction
force. Then a necessary condition is that the Hamiltonian system has a
center $(q_c,p_c)$ and a saddle point $(q_s,p_s)$ for $\eta > 0$. With
weak friction, the center becomes a focus. When $\eta$ becomes very
close to $\eta_B$ the focus becomes a node and ultimately merges with
the saddle point.

Central to the occurrence of a bifurcation is the slowing down of the
system with decreasing $\eta$. We note that this slowing down should
occur not only near the center, but in the whole range of energies
between the energies of the center $E_c=H(q_c,p_c;\eta)$ and the
saddle point $E_s=H(q_s,p_s;\eta)$. We will discuss two possible phase
portraits that meet this condition. They are shown in
Fig.~\ref{fig:scheme}.

\subsection{Local Hamiltonian bifurcation}

\begin{figure}[h]
\includegraphics[width=2.8in]{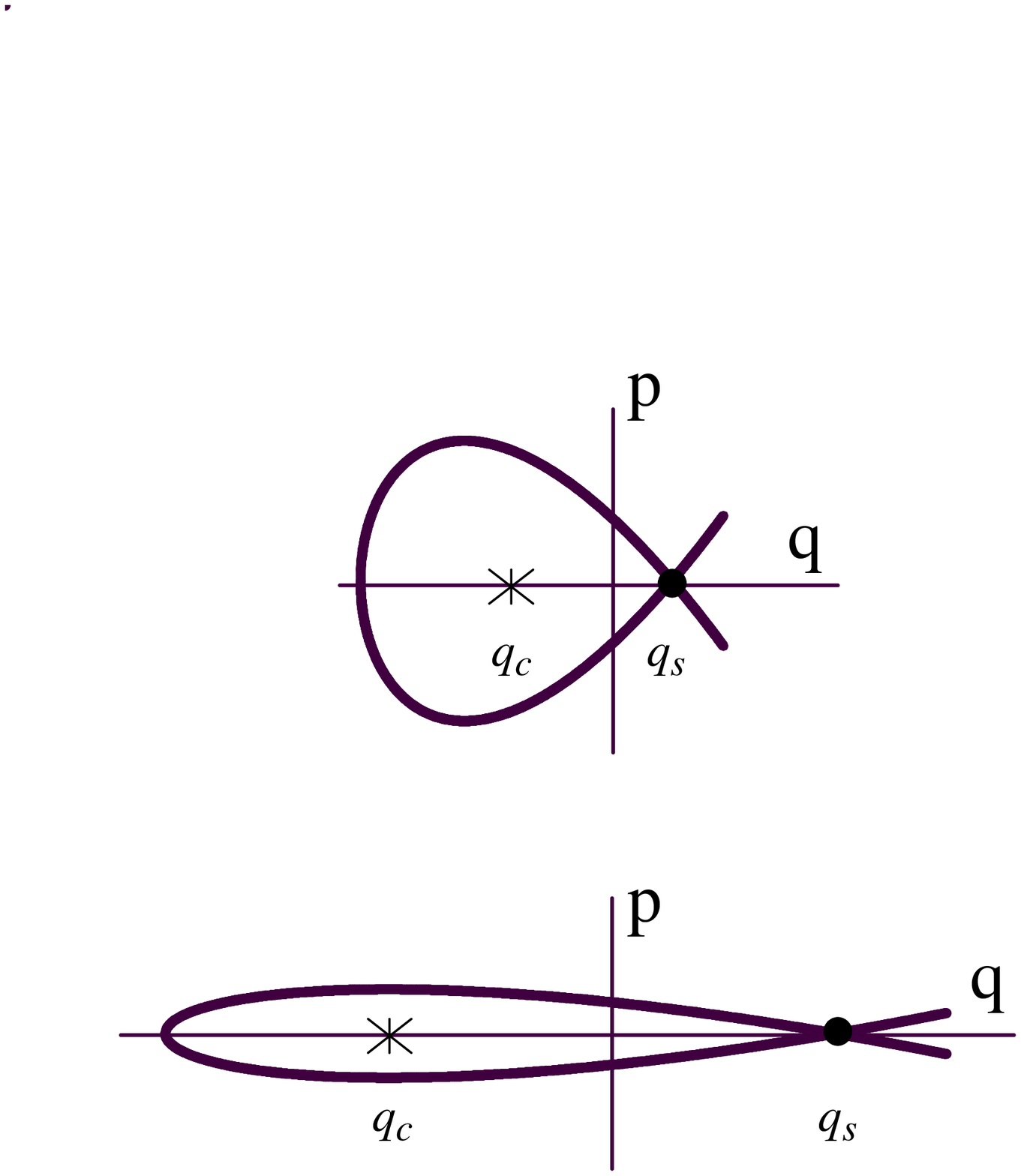}
\caption{Two types of phase portraits leading to a saddle-node
bifurcation in the presence of dissipation. The homoclinic
trajectories go through the saddle point $(q_s,p_s=0)$. The center
$(q_c,p_s=0)$ [attractor, in the presence of dissipation] is shown by the
cross. As the control parameter of the Hamiltonian $\eta$ approaches
the critical value, the loop in the left panel shrinks in all
directions. The center and the saddle approach each other in phase
space and ultimately merge at the bifurcation point. The frequency of
vibrations about the center $\omega_a$ decreases, because the center
becomes closer to the homoclinic trajectory, which has an infinite
period. In the right panel, the homoclinic orbit also shrinks as $\eta
\to 0$, but the stationary states remain separated. Still the
frequency $\omega_a$ decreases with decreasing $\eta$, because the
trajectories are pressed against the homoclinic orbit.}
\label{fig:scheme}
\end{figure}

The simplest evolution of the phase portrait of Hamiltonian dynamics
corresponds to the center and the saddle point approaching each other
in phase space with decreasing $\eta$ and ultimately merging for $\eta
=0$, see upper panel in Fig.~\ref{fig:scheme}. We call this ``local
Hamiltonian bifurcation''. The frequencies of the Hamiltonian
trajectories surrounding the center $\omega(E)$ decrease with $\eta\to
0$, because these trajectories are pressed against the homoclinic
trajectory which has zero frequency.

This case is easy to analyze. We expand the Hamiltonian $H$ in $q,p$
counted off from their value at the degenerate point into which the
center and the saddle point merge for $\eta =0$. Because of the
degeneracy, the determinant
$\partial^2_qH\partial^2_pH-(\partial_q\partial_pH)^2 =0$ at this
point. We choose $q$ along the zero-eigenvalue eigenvector of the
matrix of second derivatives of $H$ for $\eta=0$. Then to leading
order in $q,p,\eta$ the expansion of $H$ takes the form
\begin{equation}
\label{local}
H(p,q;\eta) \approx {1\over 2}p^2 + U_{\rm cub}(q)
\end{equation}
with $U_{\rm cub}(q)\equiv U_{\rm cub}(q;\eta)$ given by
Eq.~(\ref{cubic}).

In Eq.~(\ref{local}) we appropriately scaled $q,p$ and disregarded all
cubic terms except for the terms $\propto q^3$ and all terms $\propto
\eta$ except for $\eta q$. An estimate $|q|\sim \eta^{1/2}, \, |p|\sim
\eta^{3/4}, \, \omega(E)\propto \eta^{1/4}$ shows that the
higher-order terms in $q,p,\eta$ lead to small corrections for
$\eta\ll 1$. In order to make the results more intuitive, we chose the
coefficient at $p^2$ to be positive. In this case the center
$q_c=-\eta^{1/2},\, p_c=0$ has lower energy than the saddle point
$q_s=\eta^{1/2},p_s=0$. If we chose $H=-p^2/2 +U_{\rm cub}(q)$ it
would be the other way around, as we illustrate on the example below,
but this does not change the following arguments.

In the approximation (\ref{local}) the homoclinic loop that surrounds
the center has a simple form
\begin{equation}
\label{separatrix_local}
p=\pm \left({2\over 3}\eta^{3/2} - \eta q + {1\over
3}q^3\right)^{1/2}.
\end{equation}
The vibration frequency about the center is
$\omega_a=(4\eta)^{1/4}$. As expected, $\omega_a\to 0$ for
$\eta\to 0$. The difference of the energy values in the extreme
points
\begin{equation}
\label{deltaH_local}
\Delta E \equiv H(q_s,p_s;\eta)-H(q_c,p_c;\eta) =
\frac{4}{3}\eta^{3/2}.
\end{equation}

\subsection{Nonlocal Hamiltonian bifurcation}

Slowing down in the Hamiltonian system occurs also if the homoclinic
trajectory shrinks down with $\eta \to 0$ while the center $(q_c,p_c)$
and the saddle point $(q_s,p_s)$ do not move or move only slightly,
see lower panel in Fig.~\ref{fig:scheme}.  As in the local case, the
frequencies of all trajectories inside the homoclinic loop go to zero
for $\eta\to 0$. We call it a nonlocal Hamiltonian bifurcation. It
corresponds to the homoclinic orbit becoming degenerate, with no
motion along it at all, at any point. The values of the Hamiltonian at
the center and the saddle point coincide,
$H(q_c,p_c;\eta=0)=H(q_s,p_s;\eta=0)$.

We choose the momentum $p$ as the variable along which the
trajectories are shrinking for $\eta \to 0$. This variable is limited
by the ``height'' of the homoclinic loop in Fig.~\ref{fig:scheme}, and
therefore it is small for small $|\eta|$. Since on the degenerate
homoclinic trajectory $\dot q = \partial_pH =0$ for $\eta=0$, after a
proper rescaling of variables the Hamiltonian for small $\eta,p$ takes
the form
\begin{equation}
\label{nonloc_general}
H(p,q;\eta)=\frac{1}{2}p^2 + \eta U(q).
\end{equation}
A more general form
\[H=\frac{a_1^2(q)}{2}\left[p-\eta^{1/2}a_2(q)\right]^2 +\eta U(q)\]
is reduced to Eq.~(\ref{nonloc_general}) by a transition to canonical
variables $p^{\prime} = a_1(q)[p-\eta^{1/2}a_2(q)], q^{\prime}=\int
dq/a_1(q)$ (we assume that $a_1$ does not have zeros in the range of
interest).

The specific form of the potential $U(q)$ is not important, except
that $U(q)$ must have a local minimum and maximum. For $\partial^2_p
H > 0$ [as chosen in Eq.~(\ref{nonloc_general})], the
minimum of $U(q)$ corresponds to the center $q_c$, and the maximum
corresponds to the saddle point $q_s$. The homoclinic trajectory
goes around $q_c$, and therefore the turning point of this
trajectory $q_t$ (where $p=0$) lies on the opposite side from
$q_c$ with respect to $q_s$, as in Fig.~\ref{fig:scheme}. We
assume, in agreement with the qualitative picture of approaching
the bifurcation that, in the interval $[q_t,q_s]$, the potential
has no other extrema besides $q_c$ and has no singularities for
real $q$ .

A simple form of $U(q)$ consistent with the above conditions is a
cubic parabola,
%
%\begin{equation}
%\label{simplest}
$U(q)=-q^3/3 + q.$
%\end{equation}
%
In this case $q_c=-1, q_s=1$, and the turning point $q_t=-2$
independent of $\eta$. However, since the distance between $q_c$
and $q_s$ does not decrease with decreasing $\eta$, there are no
reasons to believe that the potential is generically a cubic
parabola. We will give an example of a different potential below.

For the Hamiltonian (\ref{nonloc_general}), both the momentum and
frequency scale with $\eta$ as $|p|\propto\omega(E)\propto
\eta^{1/2}$.  As expected, $\omega_a=\eta^{1/2}U''(q_c) \to 0$ for
$\eta\to 0$. To leading order in $\eta$ the energy difference at the
saddle point and the center is
\begin{eqnarray}
\label{deltaH_nonlocal}
\Delta E &\equiv& H(q_s,p_s;\eta)-H(q_c,p_c;\eta)\nonumber\\
&& =
\eta\bigl[U(q_s)-U(q_c)\bigr].
\end{eqnarray}

It follows from Eqs.~(\ref{deltaH_local}), (\ref{deltaH_nonlocal}),
that for the local and nonlocal Hamiltonian bifurcations $\Delta E$
displays scaling behavior with the control parameter, $\Delta E\propto
\eta ^{\xi}$. However, the scaling exponents $\xi$ are different,
$\xi=3/2$ and $\xi=1$ in the local and nonlocal cases, respectively.

\section{Effect of friction}

The analysis of friction is particularly simple in the case of a local
Hamiltonian bifurcation (\ref{local}). Generally, one would expect the
friction force $v^{(p,q)}$ to be smooth near the point $q=p=0$. For
small $|q|, |p|$ it can be expanded to linear terms in $q,p$. Keeping
only the terms that are not reduced to renormalization of the
parameters of the Hamiltonian, we have
\begin{equation}
\label{friction_local}
v^{(q)}=a_qq,\qquad v^{(p)}=a_pp.
\end{equation}
The center $(q_c=-\eta^{1/2}, p_c=0)$ becomes a stable state provided
$\ep(a_q+a_p) >0$. It shifts because of the friction to
$(q_a=-(\eta+\varepsilon^2a_pa_q/4)^{1/2},\, p_a= \varepsilon a_qq_a)$.
The bifurcation value of the control parameter in the local case is
\begin{equation}
\label{bif_loc_position}
\eta_B^{\rm loc}= -\varepsilon^2 a_pa_q/4.
\end{equation}

The analysis of the nonlocal case is somewhat more involved. In this
case friction changes the behavior of the system very significantly:
while the center and the saddle point remain separated for $\eta \to
0$ in the Hamiltonian case, because of the friction the corresponding
attractor and the saddle point are approaching each other as $\eta$
goes to the  bifurcation value $\eta_B^{\rm nl}$.

To understand the constraints on the friction force we note that far
from the bifurcation point, $\eta\gg \eta_B^{\rm nl}$, the motion
of the system is slowly decaying oscillations. It is characterized by
the period-average value $V_E$ of the rate of energy change
\[\dot E = -\varepsilon(v^{(p)}\partial_pH  + v^{(q)}\partial_qH). \]
It is convenient to do the averaging by writing the integral over time
as an integral over the trajectory $H(p,q;\eta)=E$, which gives
\begin{eqnarray}
\label{drift}
V_E&=&
-\varepsilon\frac{\omega(E)}{2\pi}\oint\left(dq\,
v^{(p)}-dp\,v^{(q)}\right)\nonumber\\
&&=-\varepsilon\frac{\omega(E)}{2\pi}\int\!\!\int_{A(E)} dq\,dp\,
{\bm\nabla}{\bf v},
\end{eqnarray}
where the divergence of ${\bf v}$ is defined by Eq.~(\ref{div_f}). The
circulation is taken along the trajectory $H(p,q;\eta)=E$, and $A(E)$
is the enclosed region.

We assume that the surface integral in Eq.~(\ref{drift}) does not
change sign for all energies between the center and the saddle point,
$E_c$ and $E_s$. If the integral is positive (we assume $\ep > 0$)
then the period-average energy monotonously drifts down to its value
$E_c$ if it was initially between $E_c$ and $E_s$.  This is consistent
with the assumption that friction does not lead to extra stationary
states. A sufficient and physically plausible condition is that
${\bm\nabla}{\bf v}\geq C>0$ everywhere in the area enclosed by the
homoclinic trajectory in Fig.~\ref{fig:scheme}, with $C$ independent
of $\ep, \eta$. We note that $ {\bm\nabla}{\bf v}$ must be positive at
the point $(q_a,p_a)$ in order for the state $(q_a,p_a)$ to be stable.

Close to a nonlocal Hamiltonian bifurcation point $|p|\sim
\eta^{1/2}\ll 1$, and therefore, generically, we can expand ${\bf v}(p,q)$
in $p$. The term $\propto p$ in $v^{(q)}$ can be disregarded, and it
suffices to keep only a linear in $p$ term in $v^{(p)}$; the
$p$-independent term in $v^{(p)}$ can be incorporated into the
Hamiltonian leading to a trivial renormalization $U(q)\to
U(q)+(\ep/\eta) v^{(p)}(p=0,q)$. Then the stationary states in the
presence of friction are given by the equations
\begin{eqnarray}
\label{roots_nonlocal}
&&p=\varepsilon v^{(q)},\\
&&\eta \partial_qU(q) + \varepsilon^2
\left[v^{(q)}\partial_pv^{(p)}\right]_{p=0} =0. \nonumber
\end{eqnarray}
The positions of the states remain practically unchanged with the
varying parameter of the Hamiltonian $\eta$ as long as $\eta \gg
\varepsilon^2$. When this condition holds, the frequency of vibrations
near the attractor $\omega_a\sim \eta^{1/2}$ largely exceeds their
decay rate $\propto\varepsilon$ and the motion is underdamped, as
expected.

If, as we assume, the system displays a saddle-node bifurcation for a
small $\eta$, then with decreasing $\eta$ the stationary states come
closer to each other and ultimately merge together for
$\eta=\eta_B^{\rm nl} \sim \varepsilon^2$. Operationally, the
bifurcation value of the control parameter $\eta_B^{\rm nl}$ is
determined by the condition that two solutions of
Eq.~(\ref{roots_nonlocal}) for $q,p$ coincide. An illustration for a
specific model is provided below.

We note that Eq.~(\ref{roots_nonlocal}) may have two solutions for
$\eta_B^{\rm nl}$. This means that there are two bifurcation points in
the region of small $|\eta|$. The possibility of such behavior is
clear from the Hamiltonian (\ref{nonloc_general}). For example, for
$U(q)$ of the type of a cubic parabola, the patterns of trajectories
for positive and negative $\eta$ are similar, with the center and the
hyperbolic point interchanged. The two values of $\eta_B^{\rm nl}$
correspond to the saddle-node bifurcations of these two types of
motion with decreasing $|\eta|$ in the presence of friction.

Eq.~(\ref{drift}) describes energy drift in the case of a local
bifurcation as well. In this case, from Eq.~(\ref{friction_local}),
${\bm\nabla}{\bf v}= a_p+a_q$ is constant. The integral (\ref{drift})
can be found explicitly, in terms of elliptic functions.

\section{Energy diffusion}

A major effect of noise on the dynamics of underdamped systems is
drift and diffusion of energy. Their rates are determined by the
noise intensity parameters $D_{ij}$ (\ref{white_noise}). In the
weak-noise limit, where $D_{ij}$ are the smallest parameters of
the theory, noise-induced energy drift can be disregarded compared
to the drift induced by friction. Therefore we will be interested
only in energy diffusion.

The period-average value $D_E$ of the energy diffusion coefficient can
be obtained from the equation of motion (\ref{eom}) in a standard way:
we write $dH/dt$ as $\partial_qH\dot q + \partial_pH\dot p$,
substitute the noise terms $f^{(q)}$ and $f^{(p)}$ for $\dot q$ and
$\dot p$, find the correlator of $dH/dt$ using the noise correlators
(\ref{white_noise}), and average the corresponding diffusion
coefficient over the period $2\pi/\omega(E)$.  Changing at the last
step from time integration to
integration along the trajectory $H(p,q;\eta)=E$, we obtain
\begin{widetext}
\begin{eqnarray}
\label{diffusion}
D_E&=&\frac{\omega(E)}{2\pi}\oint\left[dq\,
\bigl(D_{pp}\partial_pH + D_{pq}\partial_qH\bigr) -
dp\bigl(D_{qq}\partial_qH + D_{pq}\partial_pH\bigr)\right]
\nonumber\\
&&=\frac{\omega(E)}{2\pi}\int\!\!\int_{A(E)} dq\,dp\,
\bigl[\partial_p(D_{pp}\partial_pH + D_{pq}\partial_qH) +
\partial_q(D_{qq}\partial_qH + D_{pq}\partial_pH)\bigr]
\end{eqnarray}

Near the bifurcation point the momentum $p$ is small, $p\propto
\eta^{3/4}$ and $p\propto \eta^{1/2}$ for the local and nonlocal
bifurcation, respectively. The derivatives $\partial_qH,
\partial^2_qH$ are also small, with $\partial^2_qH\propto \eta^{1/2}$
and $\partial^2_qH\propto \eta$ for the local and nonlocal
bifurcation. Then taking into account that $\partial_pH=p$, we can
simplify Eq.~(\ref{diffusion}),
\begin{equation}
\label{diffusion_simplified}
D_E\approx \frac{\omega(E)}{2\pi}\int\!\!\int_{A(E)} dq\,dp\,
D_{pp}\partial^2_pH = \frac{\omega(E)}{2\pi}\int\!\!\int_{A(E)} dq\,dp\,D_{pp}.
\end{equation}
\end{widetext}
Here, we have assumed that the coefficient $D_{pp}$ is not
proportional to a power of a small parameter $\eta$ inside the areas
enclosed by the homoclinic orbits in Fig.~\ref{fig:scheme}.

\section{Escape rate}

Diffusion over energy leads to escape from a metastable state. The
escape rate can be calculated from the one-dimensional Fokker-Planck
equation for the energy distribution $\rho(E)$,
\begin{equation}
\label{FPE_energy}
\partial_t\rho = \partial_E\left(-V_E\rho + D_E\partial_E\rho\right)
\end{equation}
(here, again, we have disregarded the corrections $\propto D$ to the
energy drift velocity).

The escape rate $W$ is determined by the probability density to have
the energy of the saddle point (the top of the potential barrier)
$E=E_s$, given that, for small noise intensity, the system is mostly
fluctuating about the center, $E=E_c$. From Eq.~(\ref{FPE_energy}) we
have
\begin{eqnarray}
\label{rate_general}
&&W={\rm const}\times\exp(-R/D),\nonumber \\
&& R=-\int_{E_c}^{E_s} dE \, V_E\frac{D}{D_E}
\end{eqnarray}
(here, $D$ is the characteristic noise intensity, see below).
Eq.~(\ref{rate_general}) is central for the following calculation. It
immediately allows us to find the scaling of the activation energy
near a bifurcation point for the local and nonlocal bifurcation.

In the case of the local bifurcation the ratio $V_E/D_E$ can be obtained
in the explicit form. From Eq.~(\ref{friction_local}) we have
${\bm\nabla}{\bf v} =a_p+a_q$. Because the area of integration $A(E)$
in Eq.~(\ref{diffusion_simplified}) is limited to a small vicinity of
the point $q=p=0$, we can set in Eq.~(\ref{diffusion_simplified})
$D_{pp}(p,q)\approx D_{pp}(0,0)\equiv D$. Then from
Eqs.~(\ref{drift}), (\ref{diffusion_simplified}) we obtain
$V_E/D_E=-\varepsilon(a_p+a_q)/D$. The activation energy of
escape $R_{\rm loc}$ is simply proportional to the energy
difference $\Delta E=E_s-E_c$ (\ref{deltaH_local}),
\begin{equation}
\label{local_final}
R_{\rm loc}=\varepsilon(a_p+a_q)\Delta
E=\frac{4}{3}\varepsilon(a_p+a_q)\,\eta^{3/2}.
\end{equation}

Eq.~(\ref{local_final}) shows that, for the local bifurcation, in the
general case of a nonequilibrium underdamped system the activation
energy of escape scales with the distance to the bifurcation point as
$\eta^{\xi}$ with $\xi=3/2$. This is the same exponent as in the
overdamped regime.

In the case of the nonlocal Hamiltonian bifurcation the range of
integration over $p$ in the expressions (\ref{drift}),
(\ref{diffusion_simplified}) for $V_E$ and $D_E$ is narrow, $\propto
\eta^{1/2}$, whereas the range of integration over $q$ is independent
of $\eta$. Therefore the integrands ${\bm\nabla}{\bf v}$ and $D_{pp}$
in these expressions can be calculated for $p=0$. Then integration
over $p$ can be
done directly, and the integrals in Eqs.~(\ref{drift}) and
(\ref{diffusion_simplified}) take a form
\begin{eqnarray}
\label{auxiliary}
&&(E_s-E_c)^{1/2}\int_{q_{\min}(x)}^{q_{\max}(x)}dq\,f(q)\nonumber\\
&&\times\left\{x-\frac{\eta}{E_s-E_c}\left[U(q)-U(q_c)\right]\right\}^{1/2},
\end{eqnarray}
where $x\equiv x(E)=(E-E_c)/(E_s-E_c)$ is the reduced energy, and
$f(q)$ is proportional to ${\bm\nabla}{\bf v}(p=0,q)$ and
$D_{pp}(p=0,q)$,
respectively. The functions $q_{\min}$ and $q_{\max}$ determine the
limits of integration and are given by the equation $E=\eta U(q)$.

Since $\Delta E=E_s-E_c = {\rm const}\times \eta$
[cf. Eq.~(\ref{deltaH_nonlocal})], the expression (\ref{auxiliary})
has the form $(E_s-E_c)^{1/2}\tilde f(x)$, i.e., it depends on energy
only in terms of the parameter $x(E)$ and is proportional to
$\eta^{1/2}$.  Therefore the ratio $V_E/D_E$ is a function of
$(E-E_c)/(E_s-E_c)$, which is proportional to $\ep /D$, with $D$ being
the characteristic noise intensity. Then, from
Eq.~(\ref{rate_general}), to leading order in $\eta$ the
activation energy near the nonlocal bifurcation $R_{\rm nl}$ is
proportional to the energy difference $\Delta E \propto\eta$, as in
the local case,
\begin{equation}
\label{nonlocal_final}
R_{\rm nl}={\rm const}\times \varepsilon\eta.
\end{equation}

From Eq.~(\ref{nonlocal_final}) the activation energy scales with the
distance to the bifurcation point as $\eta^{\xi}$ with $\xi=1$. We are
not aware of a prediction of such dependence in the general case. It
is much slower than for the local Hamiltonian bifurcation.

\section{Example: a resonantly driven nonlinear oscillator}

A system where the both types of bifurcations may occur and which has
recently attracted significant interest
\cite{Siddiqi04,Cleland05,Mohanty_APL05,Chan05} is an underdamped
nonlinear oscillator driven by a resonant field. The motion of such an
oscillator in the rotating frame is described by Eqs.~(\ref{eom}) with
\begin{eqnarray}
\label{Duffing_eom}
H(p,q)&=&\frac{1}{4}(q^2+p^2-1)^2-\beta^{1/2}q,\nonumber\\
v^{(q)}&=&q, \qquad v^{(p)}=p.
\end{eqnarray}
For several microscopic and phenomenological models of noise, which
include those of interest for the experiment
\cite{Siddiqi04,Cleland05,Mohanty_APL05,Chan05}, the components
$f^{(q)}$, $f^{(p)}$ are independent and $\delta$-correlated, with
coordinate-independent $D_{pp}=D_{qq}=D$, $D_{pq}=0$.

The parameter $\beta$ characterizes the intensity of the driving
resonant field. For a small friction coefficient $\varepsilon$, the
oscillator is bistable in the region $\beta_B^{(1)} < \beta <
\beta_B^{(2)}$, where $\beta_B^{(1)}\approx \varepsilon^2$ and
$\beta_B^{(2)}\approx 4/27$.

\begin{figure}[h]
\includegraphics[width=3.0in]{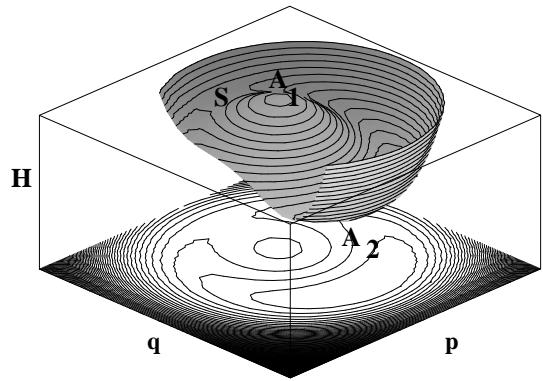}
\caption{The effective Hamiltonian $H(p,q)$ of a resonantly driven nonlinear
oscillator in the rotating frame (\protect\ref{Duffing_eom}). The plot
refers to the reduced field $\beta = 2/27$. The local maximum $A_1$
and the minimum $A_2$ are the centers, and $S$ is the saddle
point. The trajectories for constant energy $H(p,q)=E$ are shown in
the lower panel.}
\label{fig:H_function}
\end{figure}

The Hamiltonian $H$ is shown in Fig.~\ref{fig:H_function}. It has the
shape of a tilted Mexican hat. The local maximum (the top of the
internal dome) $A_1$ and the absolute minimum $A_2$ are centers; they
become attractors due to friction; $S$ is the saddle point. A general
expression for the escape rate of the oscillator was obtained in
Ref.~\onlinecite{DK79}; in particular, the activation energy was
calculated explicitly as a function of $\beta$ in the underdamped case
$\ep \ll 1$.
%later obtained in a different way by Dmitriev and Dyakonov
%\cite{Dyakonov86}.

For the bifurcation value of the control parameter $\beta =
\beta_B^{(2)}= 4/27$ in the limit of small damping, the saddle point
$S$ in Fig.~\ref{fig:H_function} merges with the dome top $A_1$ at
$q_B^{(2)}=-1/\sqrt{3}, p=0$. For $\eta= (4/27)^{1/2}- \beta^{1/2}\ll
1$ and small $\delta q= q-q_B^{(2)}, p$, to the leading order in $\delta
q,p,\eta$ the Hamiltonian (\ref{Duffing_eom}) has the form
\begin{equation}
\label{osc_localH}
H\approx -\frac{1}{3}p^2 -\frac{1}{\sqrt{3}}(\delta q)^3+
\eta(\delta q) + {\rm const}.
\end{equation}
This expression has essentially the same form as the Hamiltonian
(\ref{local}) near the local bifurcation point. Note, however, that
the center is the local {\it maximum} rather than the minimum of the
potential.

It follows from Eq.~(\ref{osc_localH}) that near the local bifurcation
point $|\Delta E|= |E_s-E_c|= (4/3^{5/4})\eta^{3/2}$. Then from
Eq.~(\ref{local_final})
\begin{equation}
\label{osc_local_result}
R_{\rm loc}\equiv R_B^{(2)}
=(4/3^{1/4})\varepsilon\eta^{3/2}=\frac{9\varepsilon}{2}
(\beta_B^{(2)}-\beta)^{3/2}
\end{equation}
(we have taken into account that, for the Hamiltonian
(\ref{osc_localH}), the last term in Eq.~(\ref{diffusion_simplified})
has an extra factor 3/2).

Eq.~(\ref{osc_local_result}) coincides with the expression obtained
for the present system by Dmitriev and Dyakonov \cite{Dyakonov86}. It
coincides also with the expression obtained from the analysis of the
oscillator dynamics very close to the bifurcation point
$\beta_B^{(2)}$, where the oscillator motion is overdamped
\cite{Dykman80}. This is similar to the situation in equilibrium
systems, where the escape rate as a function of the distance to the
bifurcation point is given by the same expression in the opposite
limits of underdamped and overdamped motion.

\begin{figure}[h]
\includegraphics[width=2.8in]{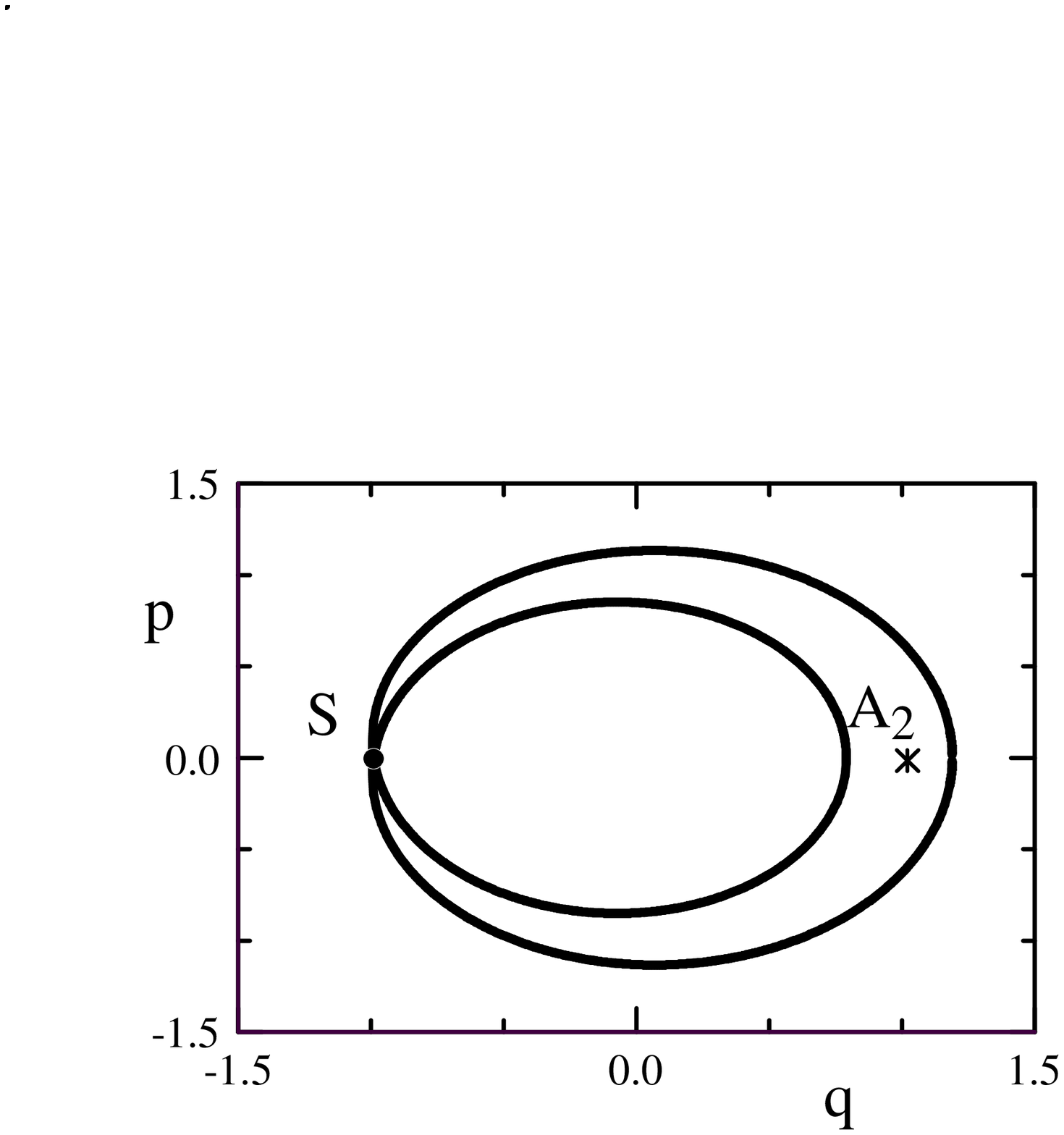}
\caption{Homoclinic trajectories of a resonantly driven oscillator
near a nonlocal Hamiltonian bifurcation.  The center $A_2$ is marked
by the star,$S$ is the saddle point. As the control parameter $\beta$
approaches the bifurcation value $\beta_B^{(1)}$ the homoclinic
trajectories approach each other and ultimately merge at the
bifurcation point. At this time the energies of the saddle and the
center become equal to each other. The data refer to $\beta=3.9\times
10^{-4}$. }
\label{fig:horseshoe}
\end{figure}

The dynamics of the oscillator near the bifurcation point
$\beta_B^{(1)}$ is described by the nonlocal bifurcation theory. The
structure of the phase portrait in the oscillator variables $(q,p)$
for the appropriate energy range is clear from
Fig.~\ref{fig:horseshoe}. The homoclinic trajectories form a double
loop with narrow space between the trajectories. The trajectories that
surround the center $A_2$ have the shape of horseshoes ``squeezed''
into the interloop space, as seen also from Fig.~\ref{fig:H_function}.

The slowing down of motion is a result of the fact that the system
moves in the opposite directions along the homoclinic
trajectories. When they approach each other (and the circle
$p^2+q^2=1$ that lies between them) with $\beta$ approaching
$\beta_B^{(1)}$, the motion is slowed down everywhere between them,
including the center.

Even though the phase portrait in Fig.~\ref{fig:horseshoe} looks
somewhat different from the one in the lower panel of
Fig.~\ref{fig:scheme}, they can be mapped onto each other. Therefore
the results for escape near a nonlocal bifurcation should apply to the
oscillator.

For $\beta \ll 1$ and $|p^2+q^2-1| \ll 1$ it is convenient to change
in Eq.~(\ref{Duffing_eom}) to new canonical variables,
$P=-(p^2+q^2-1)/2$ and $Q = \arctan(p/q)$. In these variables, to
leading order in $P, \beta$ the Hamiltonian takes the form
(\ref{nonloc_general}),
\begin{equation}
\label{osc_nonlocalH}
H=P^2 -\eta\cos Q,\qquad \eta=\beta^{1/2}.
\end{equation}
The center and the saddle point are $Q_c=0,\, P_c=0$ and $Q_s=\pi,\,
P_s=0$, respectively. The energy difference $\Delta E=E_s-E_c= 2\eta$.

From Eq.~(\ref{Duffing_eom}) and the expressions for the noise
diffusion coefficients, the corresponding functions that describe
friction force and noise in variables $Q,P$ are
\begin{eqnarray}
\label{friction_diffusion_osc}
v^{(P)}=1-2P, &&\quad v^{(Q)}=D_{PQ}=0,\nonumber \\
D_{PP}=D(1-2P),&& \quad D_{QQ}=D(1-2P)^{-1}.
\end{eqnarray}
We note that the oscillator has only one saddle-node bifurcation point
for $\beta=\eta^2 \ll 1$, because by construction $\eta > 0$.

From Eqs.~(\ref{drift}), (\ref{diffusion_simplified}),
(\ref{friction_diffusion_osc}) we obtain
$V_E/D_E=\varepsilon/D$. Therefore the activation energy of escape
near the nonlocal  bifurcation is
\begin{equation}
\label{osc_nonlocal_result}
R_{\rm nl}\equiv R_B^{(1)}
=2\varepsilon\eta=2\varepsilon
\beta^{1/2}.
\end{equation}
Eq.~(\ref{osc_nonlocal_result}) coincides with the expression in
Ref.~\onlinecite{DK79} obtained by a completely different method.

\section{Discussion of results}

We have considered activated escape in systems close to a saddle-node
bifurcation point, the problem of interest not only for recent
experiments in different fields of physics
\cite{Siddiqi04,Cleland05,Mohanty_APL05,Chan05,Schwartz_laser04}, but
also for such areas of recent attention as epidemic control
\cite{Schwartz_epidemics04}. Of special interest to us were
underdamped systems. We studied the range of the control parameter
where a system is close, but not too close to the bifurcation
point. In this range the motion is already slowed down and is
described by one degree of freedom, but is still underdamped. The
slowness of motion justifies describing it by a Langevin equation in
which delay of the friction force and finite correlation time of noise
are disregarded.

We have identified two scenarios which an underdamped system can follow as it approaches
a saddle-node bifurcation.  One is local, where in the neglect of damping the system has
a saddle point and a center close to each other in phase space. As the control parameter
$\eta$ approaches the bifurcation value $\eta_B=0$ these states approach each other. It
is the closeness of the states that leads to motion slowing down in this case.  The
second scenario is nonlocal, where the saddle point and the center remain far from each
other until damping becomes strong. However, the phase trajectories have a shape of
narrow cigars, and since on the opposite sides of the cigar the system moves in opposite
directions, the overall motion becomes slow. As $\eta\to\eta_B$, the cigars are further
squeezed, leading to further slowing down.

The major effect of friction is energy drift towards a stable state. In contrast, noise
leads to energy diffusion away from the metastable state and ultimately to escape. The
activation energy of escape $R$ is determined by the ratio of the drift and diffusion
coefficients. We found that, for both bifurcation scenarios, this ratio has a simple
form. It can be found explicitly for the local scenario, whereas for the nonlocal
scenario it has a form of a function of reduced energy. As a result, the activation
energy is proportional to the energy difference in the saddle point and the center.

We have shown that, for both types of underdamped systems, the
activation energy displays scaling $R\propto \eta^{\xi}$. The scaling
exponent $\xi=3/2$ for the local scenario and $\xi=1$ for the nonlocal
one. The scaling $\xi=3/2$ has been known for a saddle-node
bifurcation in overdamped systems and also for escape of a Brownian
particle due to thermal fluctuations. As we show, this behavior is
independent of the nature of friction and fluctuations and occurs even
in the absence of detailed balance. The scaling $\xi=1$ is a new
result. We give an example of a system of recent interest where both
types of scaling can be observed.

MID acknowledges support by the NSF DMR-0305746, IBS acknowledges
support from the Office of Naval Research, MS acknowledges support by
the NSF DMS-0401178 and BSF 2002375. The research of MID and MS was
also supported in part by the Institute for Quantum Sciences at
Michigan State University.

\end{document}